\documentclass[12pt]{article}
\newcommand{\be}{\begin{equation}}
\newcommand{\ee}{\end{equation}}
\newcommand{\bea}{\begin{eqnarray}}
\newcommand{\eea}{\end{eqnarray}}
\newcommand{\sptwo}{1.4}
\newcommand{\doublespace}{\edef\baselinestretch{\sptwo}\Large\normalsize}
\newcommand{\newsection}[1]{\section{#1}\setcounter{equation}{0}}
\renewcommand{\theequation}{\thesection.\arabic{equation}}
\newcounter{newapp}
\renewcommand{\thenewapp}{\Alph{newapp}}
\begin{document}
\vspace*{0.2in}
\begin{center}
{\large\bf $AdS_{d+1}\rightarrow AdS_d$}
\end{center}
\vspace{0.2in}
\begin{center}
{T.E. Clark}\footnote{e-mail address: clark@physics.purdue.edu}$~^a~~,~~${S.T. Love}\footnote{e-mail address: love@physics.purdue.edu}$~^{a,b}~~,~~${Muneto Nitta}\footnote{e-mail address: nitta@th.phys.titech.ac.jp}$~^c~~,~~${T. ter Veldhuis}\footnote{e-mail address: terveldhuis@macalester.edu}$~^d$\\
\end{center}
a. {\it Department of Physics, Purdue University, West Lafayette, IN 47907-1396}\\
~\\
b. {\it Fermi National Accelerator Laboratory, P.O. Box 500, Batavia, IL 60510}\\
~\\
c. {\it Department of Physics, Tokyo Institute of Technology, Tokyo 152-8551, Japan}\\
~\\
d. {\it Department of Physics \& Astronomy, Macalester College, Saint Paul, MN 55105-1899}
\vspace{0.2in}
\begin{center}
{\bf Abstract}
\end{center}
Coset methods are used to construct the action describing the dynamics of the (massive) Nambu-Goldstone scalar degree of freedom associated with the spontaneous breaking of the isometry group of $AdS_{d+1}$ space to that of an $AdS_d$ subspace.  The resulting action is an $SO(2,d)$ invariant $AdS$ generalization of the Nambu-Goto action.  The vector field theory equivalent action is also determined.
~\\

\newpage

\doublespace

\newsection{Introduction}

The $AdS/CFT$ correspondence originally proposed by Maldecena \cite{Maldecena} has sparked an enormous amount of theoretical investigation and speculation \cite{review}. The nature of this connection has been recently considered \cite{Karch:2000ct},\cite{Karch:2001cw} for the case of an $AdS_d$ defect (brane) embedded in an underlying $AdS_{d+1}$ space so that the 
$SO(2,d)$ isometry group of the $AdS_{d+1}$ metric is spontaneously broken to the $SO(2,d-1)$ isometry group of the $AdS_d$ space.  It turns out that this case also provides a surprising example of a localized D=4 gravity in which the metric differs drastically from the Minkowski metric only far from the brane \cite{Karch:2000ct}.  Moreover, when a $CFT$ is coupled to $AdS$ gravity, it has been shown that the $AdS$ graviton obtains a mass by means of the $AdS$ Higgs mechanism with the graviton absorbing a massive $AdS$ vector \cite{Dusedau:1985ue} Nambu-Goldstone 
field \cite{Karch:2000ct}\cite{Porrati:2001gx}\cite{Porrati:2001db}.

As a consequence of the embedding of the $AdS_d$ brane in $AdS_{d+1}$ space, it follows that $(d+1)$ of the  pseudo-translations and Lorentz transformations of the $AdS_{d+1}$ space are spontaneously broken.  As is typical of spontaneous space-time symmetry breaking \cite{Ivanov:1975zq}, however, not all of the broken symmetries give rise to independent Nambu-Goldstone degrees of freedom. In fact, there is only a single  Nambu-Goldstone boson, denoted $\phi (x)$,  associated with a broken pseudo-translation generator which is independent. Here $x^\mu$, $\mu= 0, 1,\ldots, d-1$, are coordinates parametrizing the $AdS_d$ space. In this paper, we construct the action governing the dynamics of this Nambu-Goldstone boson. This mode has a natural interpretation as  describing the coordinate oscillations of the brane into the target space covolume. As such we obtain the $AdS$ space generalization of the Nambu-Goto action.

$AdS_{d+1}$ space can be simply described as the $SO(2,d)$ invariant hyperboloidal hypersurface 
\be
\frac{1}{m^2} = X_0^2 -X_1^2 -X_2^2-\cdots -X_d^2 +X_{d+1}^2 = X^{\cal M} \hat\eta_{\cal MN} X^{\cal N} ,
\label{AdSd+1}
\ee
embedded in a $(d+2)$-dimensional pseudo-Euclidean space defined with invariant interval
\be
ds^2 = dX^{\cal M} \hat\eta_{\cal MN} dX^{\cal N} 
\label{interval}
\ee
characterized by the metric tensor $\hat\eta_{\cal MN}$ of signature $( +1, -1,-1,\ldots,-1,+1 )$, where ${\cal M, N} = 0,1,\ldots,d, d+1$. Here $X^{\cal M}$ are the pseudo-Euclidean space homogeneous coordinates and $m$ is a constant inverse length scale characterizing the $AdS_{d+1}$ space.

$AdS_{d+1}$ space containing an $AdS_d$ brane, which is a $d$ dimensional world volume with an $AdS$ metric embedded as a $X^d = 0$ hypersurface, can be described by the coordinates $X^{\cal M} = (X^\mu, X^d , X^{d+1})$ with
\bea
X^\mu &=& a(x^2) x^\mu \cosh{(mr)}~:~ \mu=0,1,...,d-1 \cr
X^d &=& \frac{1}{m} \sinh{(mr)} \cr
X^{d+1} &=& \frac{1}{m} b(x^2) \cosh{(mr)} .
\label{coordinates}
\eea
Here $x^\mu$ are the intrinsic coordinates of the $AdS_d$ world volume and $-\infty < r < \infty$ is the covolume coordinate. To satisfy the equation (\ref{AdSd+1}) of the $AdS_{d+1}$ hyperbola,  $a(x^2)$ and $b(x^2)$ are related as
\be
1 = m^2 x^2 a^2(x^2) + b^2(x^2) .
\label{abR}
\ee
Hence the $SO(2,d)$ invariant interval, Eq. (\ref{interval}), becomes
\be
ds^2 = e^{2A(r)} d\bar{s}^2 -dr^2 ,
\label{KR}
\ee
where the warp factor is $A(r)= \ln \cosh{(mr)}$ and $d\bar{s}^2 = dx^\mu \bar{g}_{\mu\nu}(x) dx^\nu$ is the $AdS_d$ invariant interval with $\bar{g}_{\mu\nu}$ the $AdS_d$ metric tensor.

The $AdS_d$ subspace has the isotropic coordinates $x^\mu$ of an $SO(2,d-1)$ invariant hyperboloid, $\frac{1}{m^2} = X_0^2 -X_1^2 -\cdots -X_{d-1}^2 +X_{d+1}^2$, embedded at $r=0=X^d$.  This subsurface maintains the coordinate relation Eq. (\ref{abR}).  This in turn leads to a form for the $AdS_d$ metric tensor given by
\bea
\bar{g}_{\mu\nu} (x) &=& a^2(x^2) P_{T\mu\nu}(x) + \left[ \left(a(x^2) +2x^2 \frac{da(x^2)}{dx^2} \right)^2 + 4\frac{x^2}{m^2} \left(\frac{db(x^2)}{dx^2}\right)^2\right] P_{L\mu\nu}(x) \cr
&=&a^2(x^2)P_{T\mu\nu}(x)+\frac{\left(a(x^2)+2x^2\frac{da(x^2)}{dx^2}\right)^2}{\left(1-m^2x^2a^2(x^2)\right)}P_{L\mu\nu}.
\label{gbar}
\eea
Here the transverse and longitudinal projectors for $x^\mu$ are defined as
\bea
P_{T\mu\nu}(x)&=& \eta_{\mu\nu} -\frac{x_\mu x_\nu}{x^2} \cr
P_{L\mu\nu}(x)&=&\frac{x_\mu x_\nu}{x^2} .
\eea
and $\eta_{\mu\nu}$ is the metric tensor for $d$--dimensional Minkowski space having signature $(+1,-1,\ldots,-1)$. Throughout this paper, indices are raised, lowered and contracted using $\eta_{\mu\nu}$ so that, for instance $x^2 \equiv x^\mu \eta_{\mu \nu}x^\nu$. Any use of a curved metric will be noted explicitly.  Further discussion of $AdS$ coordinates and some specfic choices for $a(x^2)$ and $b(x^2)$ are found in Appendix A.  

Since the embedding of the $AdS_d$ brane breaks the $AdS_{d+1}$ space-time symmetries to those of $AdS_d$, the associated Nambu-Goldstone bosons act as coordinates of the coset manifold corresponding to the $SO(2,d)\rightarrow SO(2,d-1)$ breakdown.  In this paper, we show, using coset 
methods \cite{Coleman:sm}\cite{Ivanov}\cite{CNtV}, that the $SO(2,d)$ invariant action governing the dynamics of the  Nambu-Goldstone boson degree of freedom takes the form 
\bea
\Gamma  &=& -\sigma \int d^d x ~\det{\bar{e}(x)}~ \cosh^d{(m\phi(x))} \sqrt{1-\frac{{\cal D}_\ell \phi(x) \eta^{\ell n}{\cal D}_n \phi(x)}{\cosh^2{(m\phi(x))}}}.
\label{NGAction1}
\eea
Here ${\cal D}_n = \bar{e}_n^{-1\mu}(x) \partial_\mu$ is the $SO(2,d-1)$ covariant derivative with $n=0,1,...,d-1$. The $AdS_d$ vielbein, $\bar{e}_\mu^{~n}$, corresponding to the metric tensor $\bar{g}_{\mu\nu} = \bar{e}_\mu^{~\ell} \eta_{\ell n} \bar{e}_\nu^{~n}$ is given by
\bea
\bar{e}_\mu^{~n}(x) &=& a(x^2) P_{T\mu}^{~~~n}(x) + \sqrt{\left[ \left(a(x^2) +2x^2 \frac{da(x^2)}{dx^2} \right)^2 + 4 \frac{x^2}{m^2} \left(\frac{db(x^2)}{dx^2}\right)^2\right]} P_{L\mu}^{~~~n}(x) \cr
&=&a^2(x^2)P_{T\mu}^{~~n}(x)+\frac{a(x^2)+2x^2\frac{da(x^2)}{dx^2}}{\sqrt{1-m^2x^2a^2(x^2)}}P_{L\mu}^{~~n}
\label{ebar}
\eea
The overall positive constant $\sigma$ (brane tension) carries mass dimension $d$. Note that the overall sign of the action is fixed so that the $\phi$ kinetic energy term is positive.  Eq. ($\ref{NGAction1}$) is the $AdS$ generalization of the Nambu-Goto action for an $AdS_d$ brane embedded in $AdS_{d+1}$ space. We see that the Nambu-Goldtone mode is massive and has non-derivative self interactions. 

The outline of the paper is as follows. The coset method construction is presented in section 2 along with the explicit form of the nonlinear realizations of the $SO(2,d)$ transformations of the Nambu-Goldstone fields.  The action building blocks secured from the Maurer-Cartan one-form also appear in this section.   Section 3 includes the transformation properties of the Maurer-Cartan one form along with the construction of the $SO(2,d)$ invariant $AdS_d$ action. In addition, this section contains the general $AdS_d$ coordinate transformations. The form of the $AdS_d$ action is shown to remain invariant under these coordinate transformations.  Finally in section 3, the vector field theory equivalent of this action is also secured. In this construction, it is the longitudnal component of the vector field which plays the role of the massive Nambu-Goldstone degree of freedom. In the Poincar\'e limit, $m\rightarrow 0$, the usual equivalence of this action to a tensor gauge theory \cite{Ha} action is obtained \cite{Clark:2004jn}.  There are two appendices. Appendix A contains a discussion of the various $AdS$ coordinate systems used in the paper, while the $AdS$ isometry charge algebra is reviewed in Appendix B. 
\pagebreak

\newsection{The Coset Construction}

The isometry group of the $AdS_{d+1}$ hyperboloid (c.f. Eq. (\ref{AdSd+1})) is $SO(2,d)$ while that of the  $AdS_d$ subspace is $SO(2,d-1)$.  The action governing the dynamics of the Nambu-Goldstone modes associated with the symmetry breakdown  $SO(2,d)\rightarrow SO(2,d-1)$ can be constructed by means of the coset method.  This technique begins by introducing the coset element $\Omega \in SO(2,d)/SO(1,d-1)$ where $SO(1,d-1)$ corresponds to the Lorentz (stability) group of transformations in $AdS_d$.  The $AdS_d = SO(2,d-1)/SO(1,d-1)$ coordinates, $y^\mu$, act as parameters for pseudo-translations in the world volume and are part of the coset so that
\be
\Omega (y) = e^{iy^\mu  P_\mu} e^{i\phi(y)Z} e^{iv^\mu (y) K_\mu},
\label{coset}
\ee
Here the symmetry generators $P_\mu$, $Z$ and $K_\mu$, as well as the Lorentz transformations generators $M_{\mu\nu}$, are defined in Appendix B.  The coset so defined corresponds to a particular choice of coordinates, specifically denoted as $y^\mu$, for the $AdS_d$ world volume.  The fields are also defined as functions of $y^\mu$.  Other coordinate choices, generically denoted by $x^\mu$, can and will be used.  In order to transform to these coordinates, the $y^\mu$ will be defined as a function of the $x^\mu$ coordinates, $y^\mu = y^\mu (x)$.  Likewise in this case, the fields will also be relabeled as functions of $x^\mu$. That is, $\phi (y) = \phi (y(x))\rightarrow \phi (x)$ and $v^m (y)= v^m (y(x))\rightarrow v^m (x)$.  Hence the coset element has the same operator structure as in equation (\ref{coset}) but now the fields and $y$ are functions of $x$ so that $\Omega (y) =\Omega (y(x))\rightarrow \Omega (x) = e^{iy^\mu (x) P_\mu} e^{i\phi (x) Z}e^{i v^\mu (x) K_\mu}$.  For the present, we use the $y^\mu$ coordinates and the fields are considered functions of $y^\mu$.  The Nambu-Goldstone field $\phi (y)$ along with $v^\mu (y)$ act as the remaining coordinates needed to parametrize the coset manifold $SO(2,d)/SO(2,d-1)$. 

Left multiplication of the coset elements $\Omega$ by an $SO(2,d)$ group element 
\be
g = e^{i\epsilon^\mu P_\mu} e^{izZ} e^{ib^\mu K_\mu} e^{\frac{i}{2} \lambda^{\mu\nu} M_{\mu\nu}},
\ee
which is specified by the (space-time independent) infinitesimal parameters $\epsilon^\mu, z, b^\mu , \lambda^{\mu\nu}$, results in transformations of the space-time coordinates and the Nambu-Goldstone fields according to the general form \cite{Coleman:sm}
\be
g\Omega = \Omega^\prime h .
\label{leftmult}
\ee
The transformed coset element, $\Omega^\prime$,  is a function of the transformed world volume coordinates and the total variations of the fields:
\be
\Omega^\prime = e^{iy^{\prime\mu}  P_\mu} e^{i\phi^\prime(y^\prime)Z} e^{iv^{\prime\mu} (y^\prime) K_\mu},
\ee 
while $h$ is a field dependent element of the stability group $SO(1,d-1)$:
\be
h= e^{\frac{i}{2} \theta^{\mu\nu}(y) M_{\mu\nu}} ,
\label{hele}
\ee
Exploiting the algebra of the $SO(2,d)$ charges displayed in Appendix B, along with extensive use of the Baker-Campbell-Hausdorf formulae, the $AdS$ transformations are obtained as
\bea
y^{\prime\mu}  &=&  \left[ 1 -z\sqrt{m^2} \tanh{\sqrt{m^2\phi^2}}\frac{\sin{\sqrt{m^2  y^2}}}{\sqrt{m^2 y^2}}\right]y^\mu  -\lambda^{\mu\nu} y_\nu \cr
 & &\qquad + \left[ P_{L}^{\mu\nu}(y) + \sqrt{m^2 y^2}\cot{\sqrt{m^2 y^2}}P_{T}^{\mu\nu}(y)\right] \epsilon_\nu \cr
 & &\qquad\qquad + \frac{\tanh{\sqrt{m^2\phi^2}}}{\sqrt{m^2}} \left[ \cos{\sqrt{m^2 y^2}}P_{L}^{\mu\nu}(y) + \frac{\sqrt{m^2 y^2}}{\sin{\sqrt{m^2 y^2}}}P_{T}^{\mu\nu}(y)\right] b_\nu \cr
 & & \cr
\phi^\prime (y^\prime) &=& \phi (y) + z\cos{\sqrt{m^2 y^2}} + b_\mu y^\mu \frac{\sin{\sqrt{m^2 y^2}}}{\sqrt{m^2 y^2}}\cr
 & & \cr
v^{\prime \mu} (y^\prime) &=& v^\mu (y) -\lambda^{\mu\nu} v_\nu -\frac{m^2}{2}\frac{\tan{\sqrt{m^2 y^2/4}}}{\sqrt{m^2 y^2/4}}(\epsilon^\mu y^\nu -\epsilon^\nu y^\mu)v_\nu \cr
 & &\qquad -z\frac{m^2}{\cosh{\sqrt{m^2\phi^2}}} \frac{\sin{\sqrt{m^2 y^2}}}{\sqrt{m^2 y^2}}\left[ P_{L}^{\mu\nu}(v) + \sqrt{v^2}\coth{\sqrt{v^2}}P_{T}^{\mu\nu}(v)\right] y_\nu \cr
 & &\qquad\qquad + \sqrt{m^2/4} \frac{\tan{\sqrt{m^2 y^2/4}}}{\sqrt{m^2 y^2/4}}\tanh{\sqrt{m^2 \phi^2}} (b^\mu y^\nu - b^\nu y^\mu ) v_\nu \cr
 & & +\frac{1}{\cosh{\sqrt{m^2\phi^2}}} \left[ P_{L}^{\mu\nu}(v) + \sqrt{v^2}\coth{\sqrt{v^2}}P_{T}^{\mu\nu}(v)\right] \cr
 & & \qquad\qquad\qquad\qquad\qquad\qquad\times\left[ \cos{\sqrt{m^2y^2}} P_{L\nu\rho}(y) + P_{T\nu\rho}(y)\right] b^\rho \cr
 & & \cr
\theta^{\mu\nu}(y) &=& \lambda^{\mu\nu} + \frac{m^2}{2} \frac{\tan{\sqrt{m^2y^2/4}}}{\sqrt{m^2y^2/4}}(\epsilon^\mu y^\nu - \epsilon^\nu y^\mu ) \cr
 & &\qquad -z\frac{m^2}{\cosh{\sqrt{m^2\phi^2}}} \frac{\sin{\sqrt{m^2 y^2}}}{\sqrt{m^2 y^2}}(v^\mu y^\nu -v^\nu y^\mu ) \frac{\tanh{\sqrt{v^2/2}}}{\sqrt{v^2}}\cr
 & &\qquad\qquad -\sqrt{m^2/4} \frac{\tan{\sqrt{m^2 y^2/4}}}{\sqrt{m^2 y^2/4}}\tanh{\sqrt{m^2 \phi^2}} (b^\mu y^\nu - b^\nu y^\mu )\cr
 & & -\frac{1}{\cosh{\sqrt{m^2\phi^2}}} \frac{\tanh{\sqrt{v^2/2}}}{\sqrt{v^2}}  \cr
 & &\qquad\qquad\qquad \times\left[ \cos{\sqrt{m^2y^2}} P_{L}^{\mu\rho}(y)   b_\rho v^\nu + P_{T}^{\mu\rho}(y) b_\rho v^\nu - (\mu \leftrightarrow \nu ) \right].\cr
 & & 
\label{variations}
\eea

The nonlinearly realized $SO(2,d)$ transformations induce a field dependent general coordinate transformation of the world volume space-time coordinates.  Using the $y^\mu$ coordinate transformation given above, the $AdS_{d+1}$ general coordinate transformation for the world volume space-time coordinate differentials is given by
\be
dy^{\prime \mu} = dy^\nu \bar{G}_\nu^{~\mu},
\label{dyprime}
\ee
with $\bar{G}_\nu^{~\mu} = \partial y^{\prime \mu}/\partial y^\nu$.  The $SO(2,d)$ invariant interval can be formed using the metric tensor ${g}_{\mu\nu}(y)$ so that $ds^2 = dy^\mu {g}_{\mu\nu}(y) dy^\nu = ds^{\prime 2} = dy^{\prime \mu} {g}^\prime_{\mu\nu}(y^\prime) dy^{\prime \nu}$ where the metric tensor transforms as 
\be
{g}^\prime_{\mu\nu} (y^\prime) = \bar{G}_\mu^{-1\rho} {g}_{\rho\sigma}(y) \bar{G}_\nu^{-1\sigma} .
\label{gprime}
\ee

The form of the vielbein (and hence the metric tensor) as well as the $SO(2,d)$ covariant derivatives of the Nambu-Goldstone boson fields and the spin connection can be extracted from the Maurer-Cartan one-form, $\Omega^{-1}d\Omega$, which can be expanded in terms of the generators as 
\be
\Omega^{-1} d\Omega = i\left[ \omega^m P_m + \omega_Z Z +\omega^m_K K_m +\frac{1}{2}\omega_M^{mn} M_{mn}\right].
\ee
Latin indices $m,n = 0,1,\ldots,d-1$, are used to distinguish tangent space (Latin indices) transformation properties from world volume (Greek indices) transformation properties. Recall all contracted indices (Latin or Greek) are summed using  the $d$-dimensional Minkowski metric tensor. Applying the Feynman formula for the variation of an exponential operator in conjunction with the Baker-Campell-Hausdorff formulae, the individual world volume one-forms appearing in the above decomposition of the Maurer-Cartan one-form are secured as 
\bea
\omega^m &=& -\frac{\sinh{\sqrt{v^2}}}{\sqrt{v^2}}v^m d\phi +\cosh{\sqrt{m^2\phi^2}}\left[ P_{vT}^{mn} + \cosh{\sqrt{v^2}}P_{vL}^{mn}\right] \bar\omega_n \cr
 & & \cr
\omega_Z &=& \cosh{\sqrt{m^2\phi^2}} \left[ d\phi - \cosh{\sqrt{m^2\phi^2}}\bar\omega_m v^m \frac{\tanh{\sqrt{v^2}}}{\sqrt{v^2}}\right]\cr
 & & \cr
\omega_K^m &=& dv^m - \left(\frac{\sinh{\sqrt{v^2}}}{\sqrt{v^2}} -1\right) \frac{[v^m v_n dv^n - v^2 dv^m]}{v^2}\cr
 & & -\bar\omega_{M}^{mn} v_n \frac{\sinh{\sqrt{v^2}}}{\sqrt{v^2}} \cr
 & & +\sqrt{m^2}\sinh{\sqrt{m^2\phi^2}}\left[ P_{vL}^{mn} + \cosh{\sqrt{v^2}}P_{vT}^{mn} \right] \bar\omega_n \cr
 & & \cr
\omega_M^{mn} &=& \bar\omega_M^{mn} +\left[ \cosh{\sqrt{v^2}} -1 \right] \frac{v^m dv^n - v^n dv^m}{v^2} \cr
 & & \cr
 & & -(\cosh{\sqrt{v^2}}-1)\left[P_{vLr}^{m}\bar\omega_{M}^{nr}-P_{vLr}^{n}\bar\omega_{M}^{mr}\right]\cr
 & & \cr
 & & +\sqrt{m^2} \sinh{\sqrt{m^2\phi^2}}\frac{\sinh{\sqrt{v^2}}}{\sqrt{v^2}}\bar\omega_r \left[ v^m P_{vT}^{rn} - v^n P_{vT}^{rm}\right] .
\label{MCOne-form}
\eea

In these expressions, the $AdS_d$ covariant coordinate differential, $\bar\omega^m$, and spin connection, $\bar\omega_M^{mn}$, are obtained from the $AdS_d$ coordinate one-form $(e^{-iy^m P_m})d (e^{iy^n P_n})=i[\bar\omega^m P_m + \frac{1}{2}\bar\omega_M^{mn} M_{mn}]$ as
\bea
\bar\omega^m &=& \frac{\sin{\sqrt{m^2y^2}}}{\sqrt{m^2y^2}} P_{T}^{mn}(y) dy_n  +P_{L}^{mn}(y) dy_n \cr
\bar\omega_M^{mn} &=& \left[ \cos{\sqrt{m^2y^2}} -1\right] \frac{(y^m dy^n -y^n dy^m)}{y^2}.
\label{AdSdoneform}
\eea
The differential $\bar\omega^m$ is related to the $y^\mu$ world volume coordinate differential via the veilbein $\bar{e}_\mu^{~m}(y)$ as
\be
\bar\omega^m = dy^\mu \bar{e}_\mu^{~m}(y) .
\ee  
Using equation (\ref{AdSdoneform}) along with $d=dy^\mu \partial_\mu^y$, this veilbein is obtained as
\be
\bar{e}_\mu^{~m}(y) = \frac{\sin{\sqrt{m^2y^2}}}{\sqrt{m^2y^2}} P_{T\mu}^{~~~m}(y)  +P_{L\mu}^{~~~m}(y) .
\label{etilde}
\ee
Thus it is seen that the $y$ coordinates correspond to the choice of embedding coordinates (c.f. Eq. (\ref{coordinates}))
\bea
a(y^2) &=& \frac{\sin{\sqrt{m^2y^2}}}{\sqrt{m^2y^2}} \cr
b(y^2) &=& \cos{(\sqrt{m^2y^2})} .
\label{ab}
\eea

The two sets of bases for coordinate differentials, $dy^\mu$ and the $AdS_{d+1}$ covariant one-form $\omega^m$, are related to each other through the veilbein $e_\mu^{~m}(y)$:
\be
\omega^m (y)= dy^\mu e_\mu^{~m}(y).
\ee
Again using that $d=dy^\mu \partial_\mu^y$, it follows from the first equality in Eq. (\ref{MCOne-form}) that the vielbein $e_\mu^{~m}$ can be factorized as
\be
e_\mu^{~m}(y) = \bar{e}_\mu^{~n}(y) {N}_n^{~m}(y) ,
\ee
where the Nambu-Goto veilbein, ${N}_n^{~m}(y)$, is 
\be
{N}_n^{~m}(y) = -v^m \frac{\sinh{\sqrt{v^2}}}{\sqrt{v^2}} {\cal D}_n \phi + \cosh{\sqrt{m^2\phi^2}}\left[ P_{Tn}^m(v) +\cosh{\sqrt{v^2}}P_{Ln}^m (v)\right],
\label{Nambu-Goto2}
\ee
and the $AdS_d$ covariant derivative is defined as ${\cal D}_n = \bar{e}_n^{-1\mu}(y) \partial_\mu^y $.  
\pagebreak

\newsection{The Invariant Action}

To construct an $SO(2,d)$ invariant action, we begin by using (c.f. Eqs. (\ref{leftmult}) and (\ref{hele})) 
\bea
\Omega^\prime(y^\prime)&=& g\Omega(y)e^{-\frac{i}{2}\theta^{mn}(y)M_{mn}}
\eea
and isolating the coefficient of $P_m$ in the decomposition of $(\Omega^{-1}d\Omega)^\prime(y^\prime)$ giving
\bea
\omega^{\prime~m}(y^\prime) P_m &=& \omega^m(y)e^{\frac{i}{2}\theta^{lr}(y)M_{lr}}P_m e^{-\frac{i}{2}\theta^{st}(y)M_{st}} .
\eea
Next, from the $SO(2,d)$ algebra (see Appendix B) it follows that
\bea
e^{\frac{i}{2}\theta^{lr}M_{lr}}P_m e^{-\frac{i}{2}\theta^{st}M_{st}}=\Lambda_m~^n (\theta) P_n ,
\eea
where $\Lambda_m^{~n} (\theta)$ is an $AdS_d$ (local) Lorentz transformation with parameter $\theta^{rs} (y)$ so that $\det{\Lambda} =1$.

Thus 
\bea
\omega^{\prime~m}(y^\prime)&=&dy^{\prime~\nu}e_\nu^{\prime~m}(y^\prime) =\omega^n(y)\Lambda_n~^m  = dy^\rho e_\rho~^n (y) \Lambda_n~^m
\eea
where $dy^{\prime~\mu}= dy^\rho \bar{G}_\rho~^\mu$, equation (\ref{dyprime}).  
Consequently the vielbeine are related as 
\bea
 e_\rho~^n (y)\Lambda_n~^m &=& \bar{G}_\rho~^\nu e_\nu^{\prime~m}(y^\prime) . 
\eea
Taking the determinant  (using $\det{\Lambda} =1$) then yields $\det{e^\prime} = (\det{\bar{G}})^{-1}~ \det{e}$.  Since the Jacobian of the $y^\mu \rightarrow y^{\prime~\mu}$ transformation is simply
\bea
d^dy^\prime &=& d^d y ~\det{\bar{G}},  
\eea
it follows that $d^dy^\prime ~\det{e^\prime} (y^\prime) = d^dy ~\det{e} (y)$.
Thus an $SO(2,d)$ invariant action is constructed as
\be
\Gamma = -\sigma \int d^d y \det{e(y)} ,
\ee
with the vacuum energy denoted by $\sigma$. Note that the action is simply the negative of the brane tension integrated over the invariant $AdS$ volume.

An additional general coordinate transformation can be made taking the $y^\mu$ coordinates to the $x^\mu$ coordinates, $y^\mu = y^\mu (x)$, so that the $SO(2,d)$ invariant interval will assume the form $ds^2 = dx^\mu g_{\mu\nu}(x) dx^\nu$ with the metric tensor given by 
\be
g_{\mu\nu}(x) = \frac{\partial y^\rho (x)}{\partial x^\mu}{g}_{\rho\sigma}(y(x))\frac{\partial y^\sigma (x)}{\partial x^\nu}= {a}_\mu^{~\rho} (x) {g}_{\rho\sigma}(y(x)) {a}_\nu^{~\sigma}(x).
\ee
The transformation matrix associated with this change of variables is defined to be ${a}_\mu^{~\rho}(x)= \frac{\partial y^\rho(x)}{\partial x^\mu}$.  Consequently the $SO(2,d)$ transformations similarly induce a field dependent general coordinate transformation
of these new world volume coordinates
\be
x^{\prime \mu} = x^\mu + {a}_\nu^{-1 \mu}(x) \delta y^\nu (x) ,
\label{xprime}
\ee
where the corresponding $SO(2,d)$ variation of $y^\mu$ is given in equation (\ref{variations}), $\delta y^\mu = y^{\prime \mu} -y^\mu$.  From equation (\ref{xprime}), the new coordinate differentials transform as
\be
dx^{\prime \mu} = dx^\nu G_\nu^{~\mu},
\ee
with $G_\nu^{~\mu} = \partial x^{\prime~\mu}/\partial x^\nu$.  Relating the coordinate differentials through the transformation matrix, $dy^\mu = dx^\nu (\partial y^\mu /\partial x^\nu )=dx^\nu {a}_\nu^{~\mu}$ and likewise for the $SO(2,d)$ transformed coordinates, $dy^{\prime \mu} = dx^{\prime \nu} (\partial y^{\prime \mu} /\partial x^{\prime\nu} )= dx^{\prime \nu} {a}_\nu^{\prime~\mu}$, the variation of the transformation matrix is obtained
\be
{a}_\nu^{\prime ~\mu} = \frac{\partial y^{\prime \mu}}{\partial x^{\prime \nu}} = \frac{\partial y^{\prime \mu}}{\partial y^{\sigma}} \frac{\partial y^{\sigma}}{\partial x^{\rho}} \frac{\partial x^{\rho}}{\partial x^{\prime \nu}}
= G_\nu^{-1 \rho} {a}_\rho^{~\sigma} \bar{G}_\sigma^{~\mu}.
\ee
Hence the invariance of the interval in terms of the new $x^\mu$ coordinates is secured since $g_{\mu\nu}^\prime (x^\prime)= G_\mu^{-1 \rho} g_{\rho\sigma}(x) G_\nu^{-1 \sigma}$.

The covariant coordinate differential one-forms are related to the new coordinate differentials $dx^\mu$ through the vielbein $e_\mu^{~m}(x)$ as
\be
\omega^m (x) = dx^\mu e_\mu^{~~m}(x) .
\ee
Thus the $y^\mu$ and the $x^\mu$ coordinate veilbeine are related by the transformation matrix
\be
e_\mu^{~m}(x) = a_\mu^{~\nu} e_\nu^{m}(y) .
\label{eaey}
\ee
In a similar fashion, the $\bar\omega^m$ one-form can also be expanded in terms of $dx^\mu$ as $\bar\omega^m = dx^\mu \bar{e}_\mu^{~m}(x) = dy^\mu \bar{e}_\mu^{~m}(y)$ thus giving
\be
\bar{e}_\mu^{~m}(x) = a_\mu^{~\nu} \bar{e}_\nu^{~m}(y).
\ee
The $AdS_d$ covariant derivative has the same form in either coordinate system,
\be
{\cal D}_m = \bar{e}_m^{-1\mu}(y) \partial_\mu^y = \bar{e}_m^{-1\mu}(y) a_\mu^{-1\nu}\partial_\nu^x = \bar{e}_m^{-1\mu}(x) \partial_\mu^x  
\ee
as does the Nambu-Goto vielbein $N_n^{~m}$. Thus Eq. (\ref{Nambu-Goto2}), after making the replacements $\phi (y) =\phi (y(x))\rightarrow \phi (x)$ and  $v^m (y) = v^m (y(x))\rightarrow v^m (x)$, reads
\be
{N}_n^{~m}(x) = -v^m \frac{\sinh{(\sqrt{v^2})}}{\sqrt{v^2}} {\cal D}_n \phi + \cosh{\sqrt{m^2\phi^2}}\left[ P_{Tn}^m (v)+\cosh{(\sqrt{v^2})}P_{Ln}^m (v)\right].
\ee
It follows that the $x$ coordinate vielbein also has the factorized form
\be
e_\mu^{~m}(x) = a_\mu^{~\nu} e_\nu^{~m}(y)= a_\mu^{~\nu} \bar{e}_\nu^{~n}(y)N_n^{~m}(y) = \bar{e}_\mu^{~n}(x) N_n^{~m}(x).
\ee
Using Eq. (\ref{eaey}), the $SO(2,d)$ invariant action in the $x$ coordinate system then becomes
\bea
\Gamma &=& -\sigma \int d^d y \det{e(y)} = -\sigma \int d^d x \det{a} ~ \det{ a^{-1}}~ \det{e(x)} \cr
 &=& -\sigma \int d^d x \det{e(x)} =-\sigma \int d^dx \det{\bar{e}(x)} ~\det{N(x)}.
\label{action2}
\eea
Finally, the determinant of the Nambu-Goto vielbein is evaluated as
\be
\det{N} = \cosh^d{(m\phi)} \cosh{(\sqrt{v^2})} \left[ 1 -\left(v^n\frac{\tanh{(\sqrt{v^2})}}{\sqrt{v^2}} \right) \left(\frac{{\cal D}_n \phi}{\cosh{(m\phi)}} \right)\right] ,
\label{detNG}
\ee
so that the action (\ref{action2}) reads
\be
\Gamma = -\sigma \int d^d x \det{\bar{e}(x)} ~ \cosh^d{(m\phi)} \cosh{(\sqrt{v^2})} 
\left[ 1 -\left(v^n\frac{\tanh{(\sqrt{v^2})}}{\sqrt{v^2}} \right) \left(\frac{{\cal D}_n \phi}{\cosh{(m\phi)}} \right)\right]
\label{action2a}
\ee

Note that the $v^n$ field is not an independent dynamical degree of freedom since it enters the action with no derivative. Hence it can be expressed in terms of the Nambu-Goldstone boson $\phi$ through its field equation as  
\be
v_n \frac{\tanh{\sqrt{v^2}}}{\sqrt{v^2}} = \frac{{\cal D}_n \phi}{\cosh{\sqrt{m^2\phi^2}}} .
\ee
This relation is identical to that obtained by setting to zero the $SO(2,d)$ invariant second equality in Eq. (\ref{MCOne-form}) from the expansion of the Maurer-Cartan one-form, $\omega_Z =0$. This is referred to as  the \lq\lq inverse Higgs mechanism" \cite{Ivanov:1975zq}. Exploiting this relation, the final form of the Nambu-Goto action for an $AdS_d$ brane embedded in $AdS_{d+1}$ target space is secured
\be
\Gamma = -\sigma \int d^d x \det{e} =  -\sigma \int d^d x \det{\bar{e}} \cosh^d{(m\phi)} \sqrt{1-\frac{{\cal D}_\ell \phi \eta^{\ell n}{\cal D}_n \phi}{\cosh^2{(m\phi)}}}.
\label{nga}
\ee
This is the $AdS$ space generalization of the Nambu-Goto action for the Nambu-Goldstone mode.  Expanding the action through terms bilinear in $\phi$ gives
\be
\Gamma =  -\sigma \int d^d x \det{\bar{e}}\left\{ 1+ \frac{1}{2}(d~m^2)\phi^2 -\frac{1}{2}\partial_\mu \phi \bar{g}^{\mu\nu} \partial_\nu \phi +\cdots \right\}.
\label{biphi}
\ee
It is seen that the Nambu-Goldstone boson carries the $(E,s)=(d,0)$ representation \cite{Dusedau:1985ue} of $SO(2,d-1)$. That is, it has mass squared equal to $d~m^2$ and hence energy $d$ in units of $m$ while being spin zero.

Note that the form of the action, Eq. ($\ref{nga}$), also follows from the invariant interval of the $AdS_{d+1}$ space, Eq. ($\ref{KR}$), provided one identifies the covolume coordinate $r$ with the Nambu-Goldstone field $\phi(x)$, $r\rightarrow \phi(x)$. Using this identification, the interval can be written as $ds^2 = dx^\mu g_{\mu\nu}(x)dx^\nu$, with 
\be
g_{\mu\nu}(x) = \bar{g}_{\mu\nu}(x) \cosh^2{(m\phi)}  -\partial_\mu \phi \partial_\nu \phi .
\ee
The $SO(2,d)$ invariant $AdS_d$ brane action, $\Gamma$, can then be constructed from $g_{\mu\nu}$ as
\be
\Gamma = -\sigma \int d^d x \sqrt{-(-1)^d \det{g_{\mu\nu}}} .
\ee
Introducing the $AdS_d$ vielbein, $\bar{e}_\mu^{~m}(x)$, via $\bar{g}_{\mu\nu} = \bar{e}_\mu^{~m} \eta_{mn} \bar{e}_\nu^{~n}$, as in Eq. (\ref{ebar}), and likewise the $AdS_{d+1}$ vielbein, $e_\mu^{~m}(x)$ as $g_{\mu\nu} = e_\mu^{~m} \eta_{mn} e_\nu^{~n}$, it then follows that $e_\mu^{~m}$ has the factorized form
\be
e_{\mu}^{~n}(x) = \bar{e}_\mu^{~\ell}(x) N_\ell^{~n}(x) ,
\ee
where the Nambu-Goto veilbein, $N_n^{~m}$ is given by
\pagebreak
\bea
N_\ell^{~n}(x) &=& \delta_\ell^{~n} \cosh{(m\phi(x))}\cr
 & & \cr
 & +&\left[ \sqrt{\left( \cosh^2{(m\phi(x))} - {\cal D}_r \phi(x) \eta^{rs} {\cal D}_s \phi(x) \right)}-\cosh{(m\phi(x))} \right] \frac{{\cal D}_\ell \phi(x) {\cal D}^n \phi(x)}{({\cal D} \phi)^2(x)} .\cr
 & & 
\eea
The action can then be expressed in terms of the vielbein as
\bea
\Gamma &=& -\sigma \int d^d x \det{e(x)} = -\sigma \int d^d x \det{\bar{e}(x)}~\det{N(x)} \cr
 &=& -\sigma \int d^d x \det{\bar{e}(x)} \cosh^d{(m\phi(x))} \sqrt{1-\frac{{\cal D}_\ell \phi(x) \eta^{\ell n}{\cal D}_n \phi(x)}{\cosh^2{(m\phi(x))}}},
\label{NGAction}
\eea
which is precisely Eq. ($\ref{nga}$).

Next reconsider the action with both Nambu-Goldstone fields $\phi$ and $v^m$ present and independent as given in  Eq. (\ref{action2a}):
\be
\Gamma = -\sigma \int d^d x \det{\bar{e}} \left[ \cosh^d{(\sqrt{m^2\phi^2})} \cosh{(\sqrt{v^2})} - \cosh^{d-1}{\sqrt{(m^2\phi^2})} \frac{\sinh{(\sqrt{v^2})}}{\sqrt{v^2}} v^m {\cal D}_m \phi \right] .
\label{phivNGaction}
\ee
Defining the ($\phi$ independent) vector density field $F^\mu$ as
\be
F^\mu \equiv \det{\bar{e}} \frac{\sinh{(\sqrt{v^2})}}{\sqrt{v^2}} v^n \bar{e}_n^{-1\mu},
\label{Fvdef}
\ee
the $v^n$ dependent terms can be expressed as $F^\mu$ and
\bea
\det{\bar{e}}~\cosh{(\sqrt{v^2})} &=& \sqrt{(\det{\bar{e}})^2 + F^\mu \bar{g}_{\mu\nu} F^\nu} \cr
 &=& \sqrt{-(-1)^{d}\det{\bar{g}_{\mu\nu}} + F^\mu \bar{g}_{\mu\nu} F^\nu} \cr
 &=& \sqrt{-(-1)^{d}\det{(\bar{g}_{\mu\nu} + F_\mu F_\nu})} ,
\eea
where the covariant vector field is 
\be
F_\mu = \frac{1}{\det{\bar{e}}}\bar{g}_{\mu\nu} F^\nu .
\ee
The action can then be written as
\be
\Gamma = -\sigma \int d^d x \left[ \sqrt{-(-1)^{d}\det{(\bar{g}_{\mu\nu} + F_\mu F_\nu})} \cosh^d{(m\phi)}  - F^\mu \partial_\mu \phi \cosh^{d-1}{(m\phi)}  \right] .
\ee
Expressing the second term on the right hand side as
\be
\partial_\mu \phi \cosh^{d-1}{(m\phi)} = \partial_\mu f(\phi) 
\ee
so that $\frac{d f}{d\phi} = \cosh^{d-1}{(m\phi)}$, the action, after integrating the second term by parts, can be written as
\be
\Gamma = -\sigma \int d^d x \left[\sqrt{-(-1)^{d}\det{(\bar{g}_{\mu\nu} + F_\mu F_\nu})} \cosh^d{(m\phi)}  + f(\phi) \partial_\mu F^\mu \right] .
\ee

Note that in the Poincar\'e limit, $m \rightarrow 0$, the action reduces to \\
$\Gamma \rightarrow -\sigma \int d^d x [\sqrt{1+ F^2}+ \phi \partial_\mu F^\mu ]$. In  that case, the $\phi$ field equation, $\delta \Gamma/\delta \phi =0=\partial_\mu F^\mu$, is just the Bianchi identity for $F^\mu$. That is, the dual of $F^\mu$ is closed. Hence $F^\mu$ can be (locally) expressed as $F^\mu = \epsilon^{\mu\nu(\rho)} \partial_\nu B_{(\rho)}$, where the $(d-2)$-form $B_{(\rho)}$ is a tensor gauge potential.  Eliminating the $\phi$ term from the action by integration by parts, the tensor gauge theory action dual to the bosonic Poincar\'e brane
Nambu-Goto action is obtained \cite{Clark:2004jn}.  The situation for $m\neq 0$ is quite different.

Now the $\phi$ field equation 
\be
\frac{\delta \Gamma}{\delta \phi} = 0 = -\sigma \frac{d f(\phi)}{d\phi} \left[ \partial_\mu F^\mu + d~m \sinh{(m\phi)} \det{\bar{e}}~ \cosh{(\sqrt{v^2})} \right] .
\label{ELphi}
\ee
can be used to eliminate it from the action. Introducing a Lagrange multiplier field $L$ to enforce this equality, the action becomes
\be
\Gamma = -\sigma \int d^d x \left[ \sqrt{-(-1)^{d}\det{(\bar{g}_{\mu\nu} + F_\mu F_\nu})} \left(T(\phi) + L ~d~ m \sinh{(m\phi)}\right) +L \partial_\mu F^\mu \right] ,
\label{LF}
\ee
where 
\be
T(\phi)= \cosh^d{(m\phi)} - d~m f(\phi) \sinh{(m\phi)} .
\ee
After employing Eq. (\ref{ELphi}) to eliminate the $\phi$ and $L$ fields, the action takes the form
\be
\Gamma = -\sigma \int d^d x \sqrt{(-1)^{d-1}\det{(\bar{g}_{\mu\nu} + F_\mu F_\nu})} \quad {T}(F) .
\label{Faction}
\ee
Here $T(\phi) = T(\phi(F))\rightarrow {T}(F)$ where the implicit dependence of $\phi$ on $F^\mu$ is given by 
\be
\sinh{m\phi} = -\frac{\partial_\mu F^\mu}{d~m \sqrt{-(-1)^{d}\det{(\bar{g}_{\mu\nu} + F_\mu F_\nu})}} ,
\ee
which follows from the equation of motion (\ref{ELphi}). 
 
Expanding $T$ through terms bilinear in the field gives
\be
{T}(F)= 1-\frac{1}{2}d~m^2 \phi^2 + {\cal O}(\phi^3) = 1-\left(\frac{1}{d~m^2  \det^2{\bar{e}}}\right) (\partial_\mu F^\mu)^2 + \cdots .
\ee
Substituting this into Eq. (\ref{Faction}) and expanding the square root, the bilinear in the field $F^\mu$ form of the action becomes
\be
\Gamma = -\sigma \int d^d x \det{\bar{e}} \left\{ 1 + \left(\frac{1}{m^2 d \det^2{\bar{e}}}\right)\left[ \frac{1}{2}(m^2 d) F^\mu \bar{g}_{\mu\nu} F^\nu - \frac{1}{2}\partial_\mu F^\mu \partial_\nu F^\nu +\cdots \right]\right\} .
\ee
The ellipses refer to the $F^\mu$ field self interactions.

The $F_\mu$ field equation then gives
\be
\det{\bar{e}} ~\bar{g}^{\mu\nu}\partial_\nu(\frac{1}{\det{\bar{e}}}\partial_\lambda F^\lambda ) -d~m^2 F^\mu = J^\mu
\label{ffe}
\ee
where $J_\mu$ contains the $F$ field self-interactions contained in the ellipses. Taking the divergence of Eq. (\ref{ffe}), it follows that the longitudnal projection $F_L\equiv \frac{1}{\det{\bar{e}}}\partial_\mu F^\mu$ satisfies the equation
\be
[\bar{g}^{\mu\nu}\partial_\mu\partial_\nu -dm^2 +\frac{1}{\det{\bar{e}}}\partial_\mu (\det{\bar{e}} ~\bar{g}^{\mu\nu})\partial_\nu ]F_L =\frac{1}{\det{\bar{e}}}\partial_\mu J^\mu .
\ee
The differential operator on the LHS acting on $F_L$ is identical to the differential operator acting on $\phi$ which appears in the $\phi$ field equation resulting from the $AdS$ Nambu-Goto action. Thus the longitudinal mode $F_L$ describes a propagating scalar degree of freedom with mass $d~m^2$ and Eq. (\ref{Faction}) is the vector field action equivalent to the Nambu-Goto action in $AdS$ space.  On the other hand, using the field equation (\ref{ffe}), the transverse component of $F$ is constrained to satisfy
\be
d~m^2~\epsilon^{\mu_1 \mu_2 \cdots \mu_d}\partial_{\mu_1}[\frac{1}{{\det{\bar{e}}}}\bar{g}_{\mu_2 \nu}F^\nu]=-\epsilon^{\mu_1 \mu_2 \cdots \mu_d}\partial_{\mu_1}[\frac{1}{{\det{\bar{e}}}}\bar{g}_{\mu_2 \nu}J^\nu] .
\ee

Indeed this equivalence can be run in reverse.  Starting with equation (\ref{Faction}), the Lagrange multiplier field $L$ can be re-introduced to give equation (\ref{LF}) (recall that $T(\phi) = \cosh^d{(m\phi)} - d~m g(\phi) \sinh{(m\phi)}$).  The fields $\phi$, $F^\mu$ and $L$ are all independent. Hence the $\phi$ field equation, $\delta \Gamma / \delta \phi =0$ implies that $L = g(\phi)$.  Substituting  this into the action along with the definition of $F^\mu$ in terms of $v^n$, equation (\ref{Fvdef}), and integrating by parts, the Nambu-Goto action equation (\ref{phivNGaction}) is once again obtained.
~\\
~\\

\noindent The work of TEC and STL was supported in part by the U.S. Department of Energy under grant DE-FG02-91ER40681 (Task B) while MN was supported by the Japan Society for the Promotion of Science under the Post-Doctoral Research Program. STL thanks the hospitality of the Fermilab theory group during his sabbatical leave while this project was completed.
\newpage

\setcounter{newapp}{1} 
\setcounter{equation}{0} 
\renewcommand{\theequation}{\thenewapp.\arabic{equation}}  

\section*{\large\bf Appendix A: \, $AdS$ Coordinates} 

An $AdS_d$ brane embedded as a $r=0$ hypersurface in an $AdS_{d+1}$ target space has intrinsic coordinates $x^\mu$ with the pseudo-Euclidean $(d+2)$-dimensional homogeneous coordinates given as in equation (\ref{coordinates})
\bea
X^\mu &=& a(x^2) x^\mu \cosh{(mr)} \cr
X^d &=& \frac{1}{m} \sinh{(mr)} \cr
X^{d+1} &=& \frac{1}{m} b(x^2) \cosh{(mr)} .
\eea
The $AdS_{d+1}$ target space and the embedded $AdS_d$ brane hyperbolic equations require the coordinate relation $1 = m^2 x^2 a^2 + b^2$. This leads to a form of the $AdS_d$ metric tensor, Eq. (\ref{gbar}), 
\bea
\bar{g}_{\mu\nu} (x) &=& a^2(x^2) P_{T\mu\nu}(x) + \left[ \left(a(x^2) +2x^2 \frac{da(x^2)}{dx^2} \right)^2 + 4\frac{x^2}{m^2} \left(\frac{db(x^2)}{dx^2}\right)^2\right] P_{L\mu\nu}(x) \cr
&=&a^2(x^2)P_{T\mu\nu}(x)+\frac{\left(a(x^2)+2x^2\frac{da(x^2)}{dx^2}\right)^2}{1-{m^2x^2a^2(x^2)}}P_{L\mu\nu}
\eea
and  the $AdS_d$ veilbein, Eq. (\ref{ebar}),
\bea
\bar{e}_\mu^{~m}(x) &=& a(x^2) P_{T\mu}^{~~~m}(x) + \sqrt{\left[ \left(a(x^2) +2x^2 \frac{da(x^2)}{dx^2} \right)^2 + 4 \frac{x^2}{m^2} \left(\frac{db(x^2)}{dx^2}\right)^2\right]} P_{L\mu}^{~~~m}(x) \cr
&=&a(x^2)P_{T\mu}^{~~~m}(x)+ \frac{\left(a(x^2) +2x^2 \frac{da(x^2)}{dx^2}\right) }{\sqrt{1-{m^2x^2 a^2(x^2)}}}P_{L\mu}^{~~~m}(x) .
\eea
Only the case where $a$ and $b$ are functions of $x^2$ are considered.

The coset construction setup naturally led to a specific choice of intrinsic coordinates denoted by $y^\mu$ for which the functions $a$ and $b$ were given by Eq. (\ref{ab})
\bea
a(y^2) &=& \frac{\sin{(\sqrt{m^2y^2})}}{\sqrt{m^2y^2}} \cr
b(y^2) &=& \cos{\sqrt{m^2y^2}} .
\eea
Hence the embedding relations for these homogeneous coordinates are
\bea
X^\mu &=&  y^\mu \frac{\sin{(\sqrt{m^2y^2})}}{\sqrt{m^2y^2}} \cosh{(mr)} \cr
X^d &=& \frac{1}{m} \sinh{(mr)} \cr
X^{d+1} &=& \frac{1}{m} \cos{(\sqrt{m^2y^2})} \cosh{(mr)} .
\eea
The metric on the $AdS_d$ brane was then given by the vielbein, Eq. (\ref{etilde}),
\be
\bar{e}_\mu^{~n}(y) = \frac{\sin{(\sqrt{m^2y^2})}}{\sqrt{m^2y^2}} P_{T\mu}^{~~~n}(y)  +P_{L\mu}^{~~~n}(y) .
\ee

A different choice of intrinsic coordinates $x^\mu$ that leaves the $AdS_d$ veilbein $\bar{e}_\mu^{~n}$ diagonal is given by
\bea
a(x^2) &=& \frac{4}{4+ m^2x^2} \cr
 & & \cr
b(x^2) &=& \left(\frac{4 -m^2 x^2}{4 +m^2x^2}\right),
\eea
hence the embedding relations for the homogeneous coordinates become
\bea
X^\mu &=&   a(x^2) x^\mu \cosh{(mr)} \cr
X^d &=& \frac{1}{m} \sinh{(mr)} \cr
X^{d+1} &=& a(x^2) \frac{1}{m} \left(\frac{4-m^2 x^2}{4}\right) \cosh{(mr)} .
\eea
The metric on the $AdS_d$ brane is then given by the vielbein
\be
\bar{e}_\mu^{~n}(x) = a(x^2) \delta_\mu^{~n} = \left(\frac{4}{4+ m^2 x^2}\right) \delta_\mu^{~m} .
\label{xebar}
\ee
Note that in the Poincar\'e limit, $m\rightarrow 0$, this reduces to the Minkowski metric and vielbein.

The transformation between the $x$ and $y$ coordinates is found by substituting the transformation $y^\mu = x^\mu f(x^2)$ into the expression for $\bar\omega^{m}$ in equation (\ref{AdSdoneform}) and requiring a diagonal vielbein. In the process, the function $a(x^2)$ is also determined. The resulting differential equation for $f$, 
\be
2 x^2 f^\prime (x^2) = -f(x^2)\left(1-\frac{\sin{\sqrt{m^2 x^2 f(x^2)}}}{\sqrt{m^2 x^2 f(x^2)}}\right),
\ee
has the solution 
\be
f(x^2) = \frac{2\tan^{-1}{\sqrt{m^2 x^2/4}}}{\sqrt{m^2 x^2}}.
\ee
Hence the coordinate transformation is given by
\be
y^\mu = x^\mu \frac{\tan^{-1}{\sqrt{m^2 x^2/4}}}{\sqrt{m^2 x^2/4}} ,
\ee
while the inverse relation is
\be
x^\mu = y^\mu \frac{\tan{\sqrt{m^2 y^2/4}}}{\sqrt{m^2 y^2/4}} .  
\ee
In the Poincar\'e limit, the $x$ and $y$ coordinates are identical, $y^\mu = x^\mu$.
In $AdS$ space the transformation matrix between these coordinates is
\be
a_\mu^{~\nu}(x) = \partial_\mu^x y^\nu (x) = \frac{\tan^{-1}{\sqrt{m^2 x^2/4}}}{\sqrt{m^2 x^2/4}} P_{T\mu}^{~~~\nu}(x) + a(x^2) P_{L\mu}^{~~~\nu}(x) .
\ee

This choice of coordinates has the advantage of simplifying the variations of the fields as well as the transformations of the coordinates themselves.  Transforming Eq. (\ref{variations}) from $y$ to $x$ coordinates yields
\bea
x^{\prime \mu} &=& x^\mu +\frac{1}{4}(4-m^2 x^2) \epsilon^\mu + \frac{m^2}{2} (\epsilon_\nu x^\nu) x^\mu - \lambda^{\mu\nu} x_\nu -z x^\mu \sqrt{m^2} \tanh{\sqrt{m^2 \phi^2}} \cr
 & & \cr
 & &~ + \frac{1}{4} \frac{\tanh{\sqrt{m^2 \phi^2}}}{\sqrt{m^2}} \left[ (4-m^2 x^2) P_{L}^{\mu\nu}(x) + (4 +m^2 x^2) P_{T}^{\mu\nu}(x) \right] b_\nu \cr
 & & \cr
\phi^\prime (x^\prime) &=& \phi (x) + z\left[ \frac{4-m^2 x^2}{4+ m^2 x^2}\right] + a(x^2) x^\mu b_\mu \cr
 & & \cr
v^{\prime \mu} (x^\prime) &=& v^\mu (x) -\lambda^{\mu\nu} v_\nu -\frac{m^2}{2} (\epsilon^\mu x^\nu -\epsilon^\nu x^\mu)v_\nu \cr
 & & \cr
 & & -z a(x^2)\frac{m^2}{\cosh{\sqrt{m^2\phi^2}}}\left[ P_{L}^{\mu\nu}(v) + \sqrt{v^2}\coth{\sqrt{v^2}}P_{T}^{\mu\nu}(v)\right] x_\nu \cr
 & & \cr
 & &\qquad\qquad\qquad+ \frac{\sqrt{m^2}}{2} \tanh{\sqrt{m^2 \phi^2}} (b^\mu x^\nu - b^\nu x^\mu ) v_\nu \cr
 & &+\frac{1}{\cosh{\sqrt{m^2\phi^2}}} \left[ P_{L}^{\mu\nu}(v) + \sqrt{v^2}\coth{\sqrt{v^2}}P_{T}^{\mu\nu}(v)\right] \cr
 & & \cr
 & &\qquad\qquad\qquad\qquad\qquad\times\left[P_{T\nu\rho}(x) + \left(\frac{4-m^2 x^2}{4+m^2 x^2}\right)P_{L\nu\rho}(x) \right] b^\rho \cr
 & & \cr
\theta^{\mu\nu} &=& \lambda^{\mu\nu} + \frac{m^2}{2} (\epsilon^\mu x^\nu - \epsilon^\nu x^\mu )\cr
 & & \cr
 & &\qquad-z a(x^2)\frac{m^2}{\cosh{\sqrt{m^2\phi^2}}} \frac{\tanh{\sqrt{v^2/2}}}{\sqrt{v^2}}(v^\mu x^\nu -v^\nu x^\mu ) \cr
 & & \cr
 & &\qquad\qquad\qquad -\frac{\sqrt{m^2}}{2} \tanh{\sqrt{m^2 \phi^2}} (b^\mu x^\nu - b^\nu x^\mu )\cr
 & & \cr
 & & -\frac{1}{\cosh{\sqrt{m^2\phi^2}}}\frac{\tanh{\sqrt{v^2/2}}}{\sqrt{v^2}}\cr
 & & \cr
 & &\qquad\qquad\times \left[  P_{T}^{\mu\rho}(x) b_\rho v^\nu + \left(\frac{4-m^2 x^2}{4+m^2 x^2}\right) P_{L}^{\mu\rho}(x)   b_\rho v^\nu + - (\mu \leftrightarrow \nu ) \right].\cr
 & & 
\label{xvariations}
\eea
\newpage

\setcounter{newapp}{2}
\setcounter{equation}{0}
\renewcommand{\theequation}{\thenewapp.\arabic{equation}}

\section*{\large\bf Appendix B: \, $AdS$ Symmetry Charge Algebra}

$AdS_{d+1}$ space can be viewed as a hyperboloid embedded in a $(d+2)$-dimensional pseudo-Euclidean space.  The equation of such a hypersurface is given in Eq. (\ref{AdSd+1}),
\be
\frac{1}{m^2} = X_0^2 -X_1^2 -X_2^2-\cdots -X_d^2 +X_{d+1}^2 = X^{\cal M} \hat\eta_{\cal MN} X^{\cal N} .
\ee
Here the pseudo-Euclidean metric tensor $\hat{\eta}_{\mu\nu}$ has signature $(+1,-1,-1,\cdots,-1,+1)$ and ${\cal {M,N}}=0,1,\cdots,d,d+1$. It follows that the the isometry group of this $AdS_{d+1}$ space is $SO(2,d)$.  Denoting the symmetry generators as $M^{\cal MN}$, they obey the algebra
\be
\left[M^{\cal MN} , M^{\cal RS} \right] = -i \left( \hat\eta^{\cal MR} M^{\cal NS} -\hat\eta^{\cal MS} M^{\cal NR} +\hat\eta^{\cal NS} M^{\cal MR} -\hat\eta^{\cal NR} M^{\cal MS} \right).
\label{SO2dAlgebra1}
\ee
The $SO(1,d)$ subgroup of Lorentz transformations in $AdS_{d+1}$ space is generated by the charges $M^{MN}$, where $M,N = 0,1,\ldots , d$.  The remaining $SO(2,d)$ generators are the pseudo-translations in $AdS_{d+1}$ space and are given by the Lorentz group vectors $P^M = m M^{d+1,M}$.  In terms of $P^M$ and $M^{MN}$, the $SO(2,d)$ algebra (\ref{SO2dAlgebra1}) reads
\bea
\left[M^{MN} , M^{RS} \right] &=& -i \left( \eta^{MR} M^{NS} -\eta^{MS} M^{NR} +\eta^{NS} M^{MR} -\eta^{NR} M^{MS} \right) \cr
\left[M^{MN} , P^{L} \right] &=& i \left( P^M \eta^{NL} - P^N \eta^{ML} \right) \cr
\left[P^M , P^{N} \right] &=& -i m^2 M^{MN} ,
\label{SO2dAlgebra2}
\eea
where the $d+1$-dimensional Minkowski metric $\eta_{MN} = (+1, -1, -1, \ldots , -1)$.

An $AdS_d$ brane embedded in the $AdS_{d+1}$ space as a $X^d = 0$ hypersurface is described by the hyperboloidal hypersurface 
\be
\frac{1}{m^2} = X_0^2 -X_1^2 -\ldots -X_{d-1}^2 +X_{d+1}^2 .
\ee
The brane spontaneously breaks the isometry group of the $AdS_{d+1}$ space from $SO(2,d)$ to $SO(2,d-1)$, which is the isometry group of the $AdS_{d}$ space.  The $SO(2,d)$ generators can be expressed in terms of the unbroken $SO(1,d-1)$ Lorentz subgroup representation content of the $SO(2,d-1)$ symmetry group of the brane.  The unbroken $SO(2,d-1)$ symmetry group is generated by the subgroup Lorentz transformation generators $M^{\mu\nu}$, where $\mu , \nu = 0,1,2,\ldots , d-1$ and the pseudo-translations in $AdS_d$ space with charges $P^\mu $.  The remaining charges are the generating elements of the $SO(2,d)/SO(2,d-1)$ coset.  They are the broken $SO(2,d)$ symmetry transformation charges.  $Z= P_{d}$ generates the broken $SO(2,d)$ pseudo-translations in the $X_d$ direction, while  $K^\mu = M^{d\mu}$ generates the broken $AdS_{d+1}$ Lorentz transformations.  Thus the $SO(2,d)$ algebra, equations (\ref{SO2dAlgebra1}) and (\ref{SO2dAlgebra2}), can be written in terms of the $P^\mu$, $M^{\mu\nu}$, $Z$ and $K^\mu$ charges as 
\bea
\left[M^{\mu\nu} , M^{\rho\sigma} \right] &=& -i \left( \eta^{\mu\rho} M^{\nu\sigma} -\eta^{\mu\sigma} M^{\nu\rho} +\eta^{\nu\sigma} M^{\mu\rho} -\eta^{\nu\rho} M^{\mu\sigma} \right) \cr
\left[M^{\mu\nu} , P^{\lambda} \right] &=& i \left( P^\mu \eta^{\nu\lambda} - P^\nu \eta^{\mu\lambda} \right) \cr
\left[M^{\mu\nu} , K^{\lambda} \right] &=& i \left( K^\mu \eta^{\nu\lambda} - K^\nu \eta^{\mu\lambda} \right) \cr
\left[M^{\mu\nu} , Z \right] &=& 0 \cr
\left[P^\mu , P^\nu \right] &=& -i m^2 M^{\mu\nu} \cr
\left[K^\mu , K^\nu \right] &=& i M^{\mu\nu} \cr
\left[P^\mu , K^\nu \right] &=& i \eta^{\mu\nu} Z \cr
\left[P^\mu , Z \right] &=& -i m^2 K^\mu \cr
\left[Z , K^\mu \right] &=& i P^\mu .
\eea
\pagebreak

\end{document}